\documentclass{eas}
\usepackage{graphicx}
\usepackage{amssymb}
\newcommand{\avg}[1]{\ensuremath{\langle \,#1\,\rangle}}
\newcommand{\Cal}[1]{\ensuremath{{\cal{#1}}}}
\begin{document}

\title{Nonlinear Structure Formation, Backreaction and Weak
  Gravitational Fields}  
\author{Aseem Paranjape}\address{Tata Inst. of Fundamental Research,
  Homi Bhabha Road, Colaba, Mumbai 400005, India.}
\runningtitle{Paranjape : Nonlinear Structure Formation \dots}
\begin{abstract}
There is an ongoing debate in the literature concerning the effects 
of averaging out inhomogeneities (``backreaction'') in cosmology. In
particular, some simple models of structure formation studied in the
literature seem to indicate that the backreaction can play a
significant role at late times, and it has
also been suggested that the standard perturbed FLRW framework is no
longer a good approximation during structure formation, when the
density contrast becomes nonlinear. In this work we use Zalaletdinov's
covariant averaging scheme (macroscopic gravity or MG) to show that
as long as the metric of the Universe can be described by the
perturbed FLRW form, the corrections due to averaging remain
negligibly small. Further, using a fully relativistic and reasonably
generic model of pressureless spherical collapse, we show that as long
as matter velocities remain small (which is true in our model), the
perturbed FLRW form of the metric can be explicitly
recovered. Together, these results imply that the backreaction remains  
small even during nonlinear structure formation, and we confirm this
within the toy model with a numerical calculation. 

\end{abstract}
\maketitle
\section{Introduction}
The application of general relativity (GR) to the large length scales
relevant in cosmology, necessarily requires an averaging operation to
be performed on the Einstein equations (Ellis \cite{ellis}). The
nonlinearity of GR then implies that such an averaging will modify the
Einstein equations on these large scales. Symbolically,  this happens
since $E[\langle g\rangle]\neq\langle E[g]\rangle$, where $g$ is the
metric and $E$ the Einstein tensor.

In recent times, attention has been focused on two promising
candidates for a consistent \emph{nonperturbative} averaging
procedure, namely the spatial averaging of scalars (Buchert
\cite{buchert}) and the covariant approach known as macroscopic
gravity (MG)(Zalaletdinov \cite{zala1,zala2}). The magnitude of the
corrections has been debated in the literature (see e.g. Kolb
\etal\ \cite{kolb}; Ishibashi \& Wald \cite{wald}; Rasanen
\cite{rasanen}; Hirata \& Seljak \cite{hirata}). Broadly speaking,
due to the structure of the Riemann tensor, the modification terms
$\cal{C}$ in any averaging approach, will have a symbolic structure
given by 
\begin{equation}
{\cal{C}} \sim \avg{{\tilde\Gamma}^2} - \avg{\tilde\Gamma}^2 \,,
\label{eq1}
\end{equation}
where $\Gamma$ denotes the Christoffel connection, and the tilde
represents any processing of the Christoffels required by the
averaging operation.

In this talk we will work within the MG framework of Zalaletdinov,
since this framework allows us to consistently define an averaged 
metric, which we will assume to be of the flat FLRW form
\begin{equation}
ds^2_{\rm FLRW} = -d\tau^2 + a(\tau)^2d\vec{x}^2\,.
\label{eq2}
\end{equation}
Specifically, we will work in the spatial averaging limit of MG, which
is relevant to cosmology (Paranjape \& Singh \cite{spatavglim}).
We will also assume, consistently with observations of the cosmic
microwave background (CMB), that the early universe was well described
by a metric which is a perturbation around the FLRW form, given by
\begin{equation}
ds^2 = -(1+2\varphi)d\tau^2 + a(\tau)^2(1-2\psi)d\vec{x}^2\,.
\label{eq3}
\end{equation}
The main argument of this talk is that although technically possible,
in the real world it is extremely unlikely that backreaction
significantly influences the average cosmological expansion.
In section 2 we will detail the
calculation of the backreaction during the linear regime of
perturbation theory, and mention some lessons that one can learn from
this calculation. In section 3 we will describe a toy model of fully
relativistic and nonlinear structure formation, and demonstrate that
the metric for this model can be explicitly brought to the perturbed
FLRW form (\ref{eq3}), \emph{even} during the nonlinear phase of the 
evolution. A calculation of the backreaction along similar lines as in
the linear case then shows that the backreaction in this toy model
remains small even at late times, thus supporting our argument. 

\section{Building the argument}
\subsection{The linear regime}
The expression in (\ref{eq1}) can be used to obtain
a simple estimate for the backreaction, assuming a perturbed FLRW
metric of the form (\ref{eq3}) with $\varphi=\psi$ in the absence of
anisotropic stresses. We then have
${\cal{C}}\sim a^{-2}\avg{\nabla\varphi\cdot \nabla\varphi}$. Assuming
a two component flat background consisting of cold dark matter (CDM)
and radiation (known as standard CDM or sCDM), and taking the
averaging to be an ensemble average over the initial conditions, it is
not difficult to show that in the matter dominated era, this estimate
leads to ${\cal{C}}\sim[10^{-4}H_0^2/a(\tau)^2]$\footnote{ The factor
  $10^{-4}$ arises from a product of the normalisation of the initial
  power spectrum $A\sim10^{-9}$, and the factor
  $(k_{eq}/H_0)^2\sim10^5$ which arises in the transfer function
  integral, where $k_{eq}=a_{eq}H(a_{eq})$ is 
  the wavenumber corresponding to the radiation-matter equality
  scale.}. This indicates that at least for epochs around the
last scattering epoch, the backreaction due to averaging was
negligible. 

The real situation is somewhat more complex than this simple
calculation indicates. On the one hand, the time evolution of
$a(\tau)$ is needed in order to solve the equations satisfied by the
perturbations, as we effectively did above by assuming a form for
$a(\tau)$.  On the other hand, the evolution of the
\emph{perturbations} is needed 
to compute the correction terms $\cal{C}$. Until these corrections are 
known, the evolution of the scale factor cannot be
determined; and until we know this evolution,
we cannot solve for the perturbations. 

To break this circle, we will adopt an iterative procedure. We first
compute a ``zeroth iteration'' estimate for the backreaction, by
assuming a fixed standard background $a^{(0)}$ such as sCDM, evolve
the perturbations and compute the time dependence of the objects
$\cal{C}$, denoted ${\cal{C}}^{(0)}$. Now, using these \emph{known}
functions of time, we form a new estimate for the background $a^{(1)}$
using the modified equations, and hence calculate the ``first
iteration'' estimate ${\cal{C}}^{(1)}$. This process can then be
repeated, and is expected to converge as long as perturbation theory
in the metric remains a valid approximation.

With these ideas in mind, we can go ahead and compute the ``zeroth
iteration'' estimate for the backreaction in the Zalaletdinov
framework. This is given by the following equations (Paranjape
\cite{pertbakrxn}) 
\begin{equation}
\left(\frac{1}{a}\frac{da}{d\tau}\right)^2 = \frac{8\pi
    G_N}{3}\bar\rho - 
\frac{1}{6}\left[ {\cal{P}}^{(1)} + {\cal{S}}^{(1)} \right]  \,,
\label{eq4}
\end{equation}
\begin{equation}
\frac{1}{a}\frac{d^2a}{d\tau^2} = -\frac{4\pi
  G_N}{3}\left(\bar\rho+3\bar p\right) + \frac{1}{3}\left[
  {\cal{P}}^{(1)} + {\cal{P}}^{(2)} +
  {\cal{S}}^{(2)} \right] \,,
\label{eq5}
\end{equation}
where we have defined 
\begin{equation}
\Cal{P}^{(1)} = -\frac{2}{a^2} \int{\frac{dk}{2\pi^2}
    k^2P_{\varphi i}(k) \left(\Phi_k^\prime \right)^2 } 
\,,
\label{eq6}
\end{equation}
\begin{equation}
\Cal{S}^{(1)} = -\frac{2}{a^2} \int{\frac{dk}{2\pi^2}
  k^2P_{\varphi i}(k) \left( k^2 \Phi_k^2\right)}
  \,,
\label{eq7}
\end{equation}
\begin{equation}
\Cal{P}^{(1)} + \Cal{P}^{(2)} = -\frac{8\Cal{H}}{a^2}
  \int{\frac{dk}{2\pi^2} k^2P_{\varphi i}(k) \left( \Phi_k\Phi_k^\prime 
    \right) } 
\,,
\label{eq8}
\end{equation}
\begin{equation}
\Cal{S}^{(2)} = -\frac{2}{a^2}   \int{\frac{dk}{2\pi^2}
  k^2P_{\varphi i}(k) \Phi_k^{\prime\prime}\left( \Phi_k -
  \frac{2\Cal{H}}{k^2}\Phi_k^\prime \right) }  
  \,.
\label{eq9}
\end{equation}
\begin{figure}[t]
\centering
\includegraphics[width=7cm]{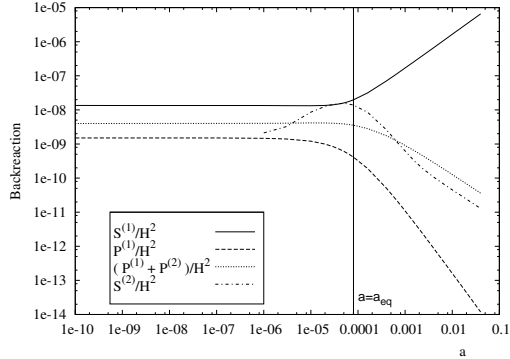}
\caption{\small The backreaction for the
  sCDM model, normalised by $H^2(a)$. $\Cal{S}^{(1)}$, $\Cal{P}^{(1)}$
  and $\Cal{S}^{(2)}$ are negative definite and their magnitudes have
  been plotted. The vertical line marks the epoch of matter radiation
  equality $a=a_{eq}$.}    
\label{fig1}
\end{figure}
Here, the prime denotes a derivative wrt. conformal time
(${}^\prime\equiv\partial_\eta = a\partial_\tau$), with
$\Cal{H}=a^\prime/a$, $\Phi_k$ is the Fourier space transfer function
defined by $\varphi(\vec{k},\eta) = \varphi_{i\vec{k}}\Phi_k(\eta)$,
and $P_{\varphi i}(k)$ is the initial power spectrum of $\varphi$. The
results of a numerical calculation 
are shown in figure \ref{fig1}, where all functions are normalised by
the Hubble parameter $H^2(a)$, and confirm that this zeroth iteration
estimate in fact gives a negligible contribution. Further, were we to
carry out the next iteration, we would essentially obtain no
difference between the zeroth and first iteration scale factors upto
the accuracy of the calculation, and hence this iteration has
effectively converged at the first step itself.

\subsection{Lessons from linear theory}
We see that the dominant contribution to the backreaction at late
times, is due to a
curvature-like term $\sim a^{-2}$, as expected from our simple
estimate above. In order to obtain a correction which grows faster
than this, we need a nonstandard evolution of the metric potential 
$\varphi$, which can only happen if the \emph{scale factor} evolves
very differently from the sCDM model, which in turn would require a
significant contribution from the backreaction. The same circle of
dependencies as before, now implies that \emph{as long as the metric 
  is perturbed FLRW}, the backreaction appears to be dynamically
suppressed.  

\begin{figure}[t]
\centering
\includegraphics[width=7cm]{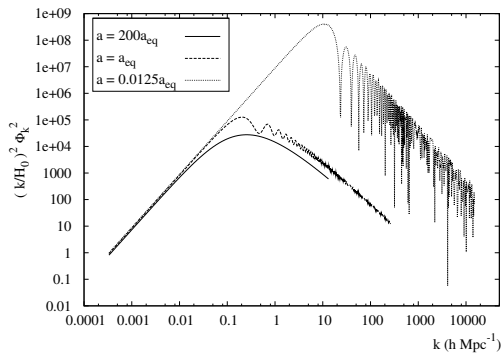}
\caption{\small The dimensionless integrand of $\Cal{S}^{(1)}$, namely
the function $(k/H_0)^2\Phi_k^2$, at three sample values of the scale
factor. The function dies down rapidly for large $k$, with the value
at some $k$ being progressively smaller with increasing scale
factor. The declining behaviour of the curves for $a=a_{eq}$ and
$a=200a_{eq}$ extrapolates to large $k$.}  
\label{fig2}
\end{figure}

\begin{figure}[t]
\centering
\includegraphics[width=7cm]{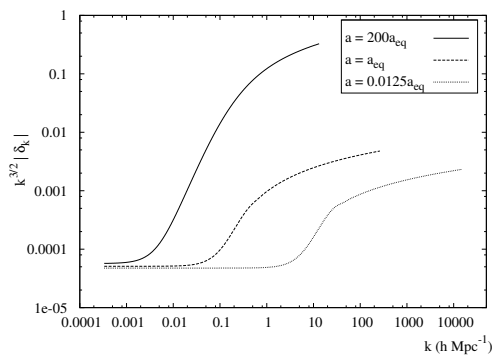}
\caption{\small The dimensionless CDM density contrast. Together with
  figure \ref{fig3} this shows that nonlinear scales do not impact the 
  backreaction integrals significantly. }   
\label{fig3}
\end{figure}

Secondly, as figures \ref{fig2} and \ref{fig3} show, scales which are
approaching nonlinearity, \emph{do not} contribute significantly to
the backreaction, which is a consequence of the suppression of small
scale power by the transfer function. We will return to this point
when discussing the backreaction during epochs of nonlinear structure
formation. 

\section{The nonlinear regime}
Let us now ask whether one can make meaningful statements concerning
the backreaction during epochs of nonlinear structure formation, when
matter density contrasts become very large and perturbation theory in
the matter variables has broken down. We begin by considering some
order of magnitude estimates.
\subsection{Dimensional arguments, and why they fail}
Let us start with the assumption that although the matter
perturbations are large, one can still expand the \emph{metric} as a
perturbation around FLRW. We are looking for either self-consistent
solutions using this assumption, or any indication that this
assumption is not valid. Given that the metric has the form
(\ref{eq3}) (and further assuming $\varphi=\psi$ as before), the
relevant gravitational equation at late times and at length scales
small comparable to $H^{-1}$, is the Poisson equation given by
\begin{equation}
\frac{1}{a^2}\nabla^2\varphi = 4\pi G\bar\rho\delta\,,
\label{eq10}
\end{equation}
where $\delta\equiv (\rho(t,\vec{x})/\bar\rho(t) - 1)$ is the density
contrast of CDM\footnote{We are only worried about the small,
  sub-Hubble scales, since larger scales are well described by linear
  theory where we know the form of the backreaction.}. As before, we
can estimate the dominant backreaction component to be ${\cal{C}}\sim 
a^{-2}\avg{\nabla\varphi\cdot \nabla\varphi}$.

Now, for an over/under-density of physical size $R$, treating
$a^{-1}\nabla\sim R^{-1}$ and $G\bar\rho\sim H^2$ on dimensional
grounds, we have
\begin{equation}
\left|\varphi\right| \sim (HR)^2\left|\delta\right|\,.
\label{eq11}
\end{equation}
For voids, we can set $\delta\sim-1$, and then $\Cal{C}\sim H^2(HR)^2\ll
H^2$, since we have assumed $HR\ll1$\footnote{I thank Karel Van
  Acoleyen for pointing this out to me.} . This shows that sub-Hubble 
underdense voids are expected to give a negligible backreaction. For
overdense regions we need to be more careful, since here $\delta$ can
grow very large. In a typical spherical collapse scenario, the
following relations hold,
\begin{equation}
R\sim(1-\cos u)r~;~ H^{-1}\sim (G\bar\rho)^{-1/2}\sim  t\sim
H_0^{-1}(u-\sin u)\,,
\label{eq12}
\end{equation}
\begin{equation}
G\rho\sim \frac{(H_0r)^2}{R^2R^\prime}\sim \frac{H_0^2}{(1-\cos u)^3}
~;~ \delta \sim (\rho/\bar\rho) \sim \frac{(u-\sin u)^2}{(1-\cos u)^3}
\,, 
\label{eq13}
\end{equation}
which lead to
\begin{equation}
\left|\varphi\right| \sim \frac{(H_0r)^2}{(1-\cos u)} ~~;~~ 
\Cal{C}\sim H^2\left[(H_0r)^2\frac{(u-\sin u)^2}{(1-\cos u)^4}  
  \right]\,. 
\label{eq14}
\end{equation}
It would therefore appear that at late enough times, the perturbative
expansion in the metric breaks down with $|\varphi|\sim1$, and the
backreaction grows large $|\Cal{C}|\sim1$. However, the crucial
question one needs to answer is the following : Is this situation
actually realised in the universe, or are we simply taking these
models too far? We make the claim that perturbation theory in
the metric \emph{does not} break down at late times, since
\emph{observed peculiar velocities remain small}. The spherical
collapse model is not a good approximation when \emph{model} peculiar 
velocities in the collapsing phase grow large. To support this claim,
we will work with an exact toy model of spherical collapse.
\begin{figure}
\centering
\includegraphics[width=7cm]{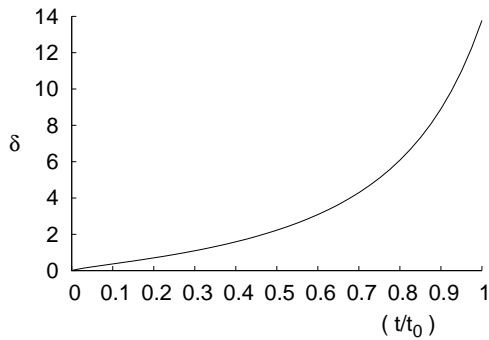}
\caption{Density contrast in the LTB toy model, at $r=8.35\,$Mpc. $t_0
  = 2/3H_0 \approx9$ Gyr.}  
\label{fig4}
\end{figure}

\subsection{Calculations in an exact model}
The model we consider was presented by Paranjape \& Singh
(\cite{sphcoll}), and can be summarized as follows. The matter content 
of the model is spherically symmetric pressureless ``dust'', and hence
the relevant exact solution is the Lema\^itre-Tolman-Bondi (LTB)
metric given by 
\begin{equation}
ds^2 = -dt^2 + \frac{R^{\prime2}dr^2}{1-k(r)r^2} + R^2d\Omega^2 \,.
\label{eq15}
\end{equation}
Here $t$ is the proper time measured by observers with fixed
coordinate $r$, which is comoving with the dust. $R(t,r)$ is the
physical area radius of the dust shell labelled by $r$, and satisfies
the equation ${\dot R}^2 = 2GM(r)/R - k(r)r^2$.
Here $M(r)$ is the mass contained inside each comoving shell, and a
dot denotes a derivative with respect to the proper time $t$. The
energy density of dust measured by an observer comoving with it
satisfies the equation $\rho(t,r) = M^\prime(r)/4\pi R^2R^\prime$, 
where the prime now denotes a derivative with respect to the LTB
radius $r$.

Initial conditions are set at a scale factor value of $a_i=10^{-3}$,
and are chosen such that the initial situation describes an FLRW
expansion with a perturbative central overdensity out to radius
$r=r_\ast$, surrounded by a perturbative underdensity out to radius
$r=r_v$, with appropriately chosen values for the various parameters
in the model (see Table 1 of Paranjape \& Singh
\cite{sphcoll}). Figure \ref{fig4} shows the evolution of the
overdensity contrast in the central region. Clearly, at late times the
situation is completely nonlinear.
Nevertheless, it can be shown that a coordinate transformation in this
model can bring its metric to the form (\ref{eq3}), \emph{provided}
one has $\left|a{v}\right| \ll1$ where ${v} = \partial_t(R/a)$. 
Physically ${v}$ is the ``comoving'' peculiar velocity. The metric
potentials (which are actually equal at the leading order, see Van
Acoleyen \cite{karel}) have the
expressions $\varphi = -\dot\xi^0 + (1/2)(a{v})^2$, $\psi =
\xi^0H + \xi$, where $\xi$ and $\xi^0$ are obtained by integrating 
$\xi^\prime = (1/2)(k(r)r^2+(a{v})^2)(R^\prime/R)$ and $\xi^{0\prime}
= a{v}R^\prime$.
A numerical calculation shows that $av$ and hence the metric
potentials do in fact remain small for the entire evolution, for
\emph{this} model. Further, the infall peculiar velocity can only
become large if the true infall velocity $\dot R$ is large,
in which case the specific background chosen to define the peculiar
velocity, becomes irrelevant (since $HR\ll1$). Hence, the fact that
relativistic infall velocities are \emph{not} observed in real
clusters etc., leads us to expect very generally that the perturbed
FLRW form for the metric should in fact be recoverable even at late
times. 
\begin{figure}[t]
\centering
\includegraphics[width=7cm]{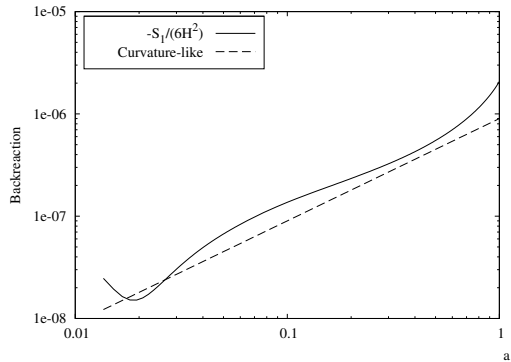}
\caption{\small The evolution of $|\Cal{S}^{(1)}|/6H^2$. Also shown is
  a hypothetical curvature-like correction, evolving like $\sim
  a^{-2}$.}    
\label{fig5}
\end{figure}

Finally, figure \ref{fig5} shows the dominant contribution to the
backreaction in the toy model (Paranjape \& Singh
\cite{nlmgcoll}). There is a significant departure from a 
curvature-like behaviour, due to evolution of the metric
potentials. More importantly, the maximum value of the backreaction 
here is $\sim10^{-6}H^2$, as opposed to $\sim10^{-4}H^2$ as seen in
the linear theory. This can be understood by noting that the
inhomogeneity of our toy model is only on relatively small, nonlinear
scales ($\lesssim20h^{-1}$Mpc), and the value of the backreaction is
therefore consistent with our earlier observation that nonlinear
scales contribute negligibly to the total backreaction. To conclude,
we have seen that as long as the metric has the perturbed FLRW form,
the backreaction remains small. Further, there are strong reasons to
expect that the metric remains a perturbation around FLRW even at late
times during nonlinear structure formation, a claim that is supported
by our toy model calculation. It should be possible to test this claim
in $N$-body simulations as well. It appears therefore, that
backreaction cannot explain the observed acceleration of the universe.  

I thank the organisers, especially Thomas Buchert, for their kind
hospitality and for holding a very lively conference in the
beautiful city of Lyon.


\end{document}